\documentstyle[11pt,psfig]{article}
\psfigurepath{/home/usuarios/fernando/PAPERS/TSALLIS/FIG}

\begin{document}

{\Large
{\bf Intermittency and Nonextensivity in Turbulence} \\
{\bf and Financial Markets}
}
\bigskip
        
F. M. Ramos\footnote{e-mail:fernando@lac.inpe.br}, C. Rodrigues Neto and R. R. Rosa
\smallskip

{\small

Laborat\'orio Associado de Computa\c c\~ao e Matem\'atica Aplicada (LAC) \\
Instituto Nacional de Pesquisas Espaciais (INPE) \\
S\~ao Jos\'e dos Campos - SP, Brazil 

\bigskip

PACS.02.50-r - Probability theory, stochastic processes, and statistics. \\
PACS.47.27Eq - Turbulence simulation and modeling. \\
PACS.89.90+n - Other areas of general interest to physicists.

}
\bigskip


{\small{\bf Abstract.} - We present a new framework for modeling the statistical 
behavior of both fully developed turbulence and short-term dynamics of financial 
markets based on the nonextensive thermostatistics proposed 
by Tsallis. We also show that intermittency -- strong bursts in the energy 
dissipation or clusters of high price volatility -- and nonextensivity -- 
anomalous scaling of usually extensive properties like entropy -- are
naturally linked by a single parameter $q$, from the
nonextensive thermostatistics.}

\bigskip

Scaling invariance plays a fundamental role in many natural 
phenomena and frequently emerges from some sort of underlying 
cascade process. A classical example is fully developed homogeneous 
isotropic three-dimensional turbulence, which is characterized by a 
cascade of kinetic energy from large forcing scales to smaller and 
smaller ones through a hierarchy of eddies. At the end
of the cascade, the energy dissipates by viscosity, turning into heat.
Recently, some authors$^{1-3}$ have studied the phenomenological
relationship among financial market dynamics, scaling behavior
and  hydrodynamic turbulence. Particularly, Ghashghaie {\it et al.}$^{2}$ 
conjectured the existence of a temporal information cascade similar 
to the spatial energy cascade found in fully developed turbulence.  

Traditionally, the properties of turbulent flows 
are studied from the statistics of velocity differences 
$v_r(x) = v(x) - v(x+r)$ at different scales $r$. As with other
physical systems that depend on the dynamical evolution of a large number
of nonlinearly coupled subsystems, the energy cascade in turbulence
generates a spatial scaling behavior -- power-law behavior with $r$ --
of the moments $\langle v_r^n \rangle$ of the probability distribution 
function (PDF) of $v_r$ (the angle brackets $\langle \rangle$ denote
the mean value of the enclosed quantity). For large values of the Reynolds number, which measures 
the ratio of nonlinear inertial forces to the linear dissipative 
forces within the fluid, there is a wide separation between the scale 
of energy input 
(integral scale $L$) and the viscous dissipation scale (Kolmogorov 
scale $\eta$). Though at large scales ($\sim L$) the PDFs are 
normally distributed, far from the integral scale they are strongly 
non-Gaussian and display wings fatter than expected for a normal process. 
This is the striking signature of the intermittency phenomenon. 
After publication of the Kolmogorov K62 refined similarity hypotheses$^{8}$,
the problem of small scale intermittency became one of the
central questions on isotropic turbulence. Over the past years 
several papers$^{9-16}$ have discussed intermittency and 
the so-called `PDF problem'. Similar 
attempts$^{1,2,17}$ have been made to explain the same peculiar shape observed in the  
PDF of price changes $z_{\tau} = z(t) - z(t + \tau)$ at small time intervals.

Based on the scaling properties of multifractals, Tsallis$^{4-7}$ has 
proposed a generalization of Boltzmann-Gibbs thermostatistics
by introducing a family of generalized nonextensive entropy functionals 
$S_q[p]$ with a single parameter $q$. These functionals reduce to the classical, 
extensive Boltzmann-Gibbs form as $q \rightarrow 1$. Optimizing $S_q[p]$ subject to appropriate
constraints$^{6}$, we obtain the distribution
\begin{equation}
\label{pdf}
p_q(x) = [1 - \beta (1-q) x^2]^{1/(1-q)}/Z_q~~.
\end{equation}
The appropriate normalization factor, for $1 < q < 3$, is given by 
\[
Z_q \equiv \left[ \frac{\beta (q-1)}{\pi} \right]^{1/2} 
\frac{\Gamma(1/(q-1))}{\Gamma((3-q)/2(q-1))}~~. 
\]
In the limit of $q \rightarrow 1$, we recover the Gaussian distribution.  

The above distribution, we claim, provides the simplest and most accurate model
for handling the PDF problem. To show this, we stay in the context of fully 
developed turbulence  ($x \equiv v_r$). From equation (\ref{pdf}), we can  
easily obtain the second moment 
\begin{equation}
\label{sec}
\langle v_r^2 \rangle = \frac{1}{\beta (5 - 3 q)}~~,
\end{equation}
and the flatness coefficient (kurtosis)
\begin{equation}
\label{kur}
K_r = \frac{\langle v_r^4 \rangle}{\langle v_r^2 \rangle^2} 
= \frac{3 \, (5 - 3 q)}{(7 -  5 q)}~~.
\end{equation}
We remark that the flatness coefficient, which is directly related to 
the occurrence of intermittency, is solely determined by the parameter $q$.
Also, we note that the positiveness of $K_r$ sets an upper bound on the 
value of $q$, namely $q < 7/5$. This bound coincides with the one obtained 
by Boghossian$^{18}$ through a $q$-generalization of 
Navier-Stokes equations. Moreover, this limit implies that the second moment
of distribution (\ref{pdf}) will always remain finite, which is empirically 
expected from the phenomena here analyzed. 

At this point, if we assume$^{2,9,10}$ a scaling of the moments 
$\langle v_r^n \rangle$ of $v_r$ as $r^{\zeta_n}$, 
the variation with $r$ of the PDF of the velocity 
differences and of its related moments can be completely determined. 
Under the assumptions of Kolmogorov log-normal model$^{8}$, 
$\zeta_n = n/3 - \frac{1}{18} \mu n (n-3)$, where $\mu$ is the 
intermittency exponent.
Then, we immediately obtain the functional forms of the flatness coefficient 
and the parameter $q$, respectively
\begin{equation}
\label{kur2}
K_r = K_L (r/L)^{\alpha} 
\end{equation} 
and
\begin{equation}
\label{qq}
q =\frac{5 - 7 (r/L)^{\alpha}}{3 - 5 (r/L)^{\alpha}}~~,
\end{equation}
with $K_L =3$, the expected value for a Gaussian process, and 
$\alpha = - 4 \mu/9$.
The correspondent expression for $\beta$ can be derived similarly from 
equation (\ref{sec}). However, to account for the well known asymmetry of the 
velocity distributions we may consider $\beta=\beta^+$, for $v_r \geq 0$, 
and $\beta=\beta^-$, for $v_r < 0$. In this case, both the second and third 
moments of the modified PDF shall be used to determine $\beta^-$ and $\beta^+$.
Equations (\ref{kur}) and (\ref{qq}) remain unchanged, as far as the asymmetry 
of the PDF is small.

We checked our model with turbulence statistics data taken from 
reference 2, provided by Chabaud {\it et al.}$^{13}$. 
Firstly, we adjusted by least-squares to the experimental data, the 
parameters of our PDF,
considering both the symmetric and asymmetric forms of equation (\ref{pdf}).
Then, we compared the estimated values of $q$ with the predictions of equation (\ref{qq}), 
for a intermittency coefficient of $\mu =0.25$ (the best estimate currently 
available is$^{12}$ $0.25 \pm 0.05$). The results are displayed in Figs. 1a and 2. 
A good agreement is observed through all spatial scales and for all orders of normalized
velocity differences. Note that the solid line in Fig. 2 is not simply a fit 
to the data, since $\mu$ is obtained independently. We also computed value of $q$ 
at the Kolmogorov scale and obtained $q_{\eta} \simeq 9/7$.  

The same approach adopted in turbulence can be 
straightforwardly applied (with $x \equiv z_{\tau}$) to model the statistics 
of price differences in 
financial markets, as far as the relevant parameters at the integral time 
scale -- time span for which a convergence to a Gaussian process is found --
are available. We tested our model with price changes data taken from 
reference 2, provided by Olsen \& Associates. The results are displayed 
in Figs. 1b and 2. Since we do not have an independent 
estimate of the intermittency coefficient of the information cascade, 
we also adjusted equation (\ref{qq}) to the data and obtained 
$\alpha = -0.17$ and $\tau_L \simeq 2.2$ days, the corresponding integral scale 
of the process. We observe that the proposed model reproduces with good
accuracy the statistics of price differences over all temporal scales. 
However, we remark that the integral scale value of 2.2 days strongly 
disagrees with other estimates$^{1}$ of $\tau_L$ (roughly 1 month).

Nonextensivity, a matter of speculation in some areas$^{19}$, is an essential 
feature of Tsallis generalized thermostatistics. If we suppose a
scenario of a cascade of bifurcations with $n$ levels, and scale 
the generalized entropy $S_q$, averaged over a volume of size 
$V=\eta^3$ and normalized by Boltzmann constant, we have at the first
level 
\begin{equation}
S_q(2V) = 2 S_q(V) + (1-q) S_q^2(V)
\end{equation}
and at the top of the cascade
\begin{equation}
\label{casca1}
S_q(2^n V) = 2 S_q(2^{n-1}V) + (1-q) S_q^2(2^{n-1}V)
\simeq 2^n S_q(V) + (1-q) 2^{n-1} S_q(V)~~,
\end{equation}
where the higher order terms in $S_q$ have been discharged. Cascade processes
are also described in terms of fractal or multifractals models$^{11,20-23}$. 
Within these frameworks, in high Reynolds number turbulence, the energy
dissipation is not uniformly distributed within the fluid but rather
concentrated on subsets of non-integer fractal $D_F$ dimension. This picture 
leads to a scaling behavior 
with dimensionality not equal to that of the embedding space. 
In this case, if we consider the cascade of bifurcations described above, we find
\begin{equation}
\label{casca2}
S_q(2^n V) = 2^{n D_F/3} S_q(V)~~.
\end{equation}
It follows immediately from equations (\ref{casca1}) and (\ref{casca2}) that
\begin{equation}
\label{casca3}
D_F \simeq \frac{3}{n} \left[ \frac{log(3 - q)}{log(2)} + n - 1 \right]~~.
\end{equation}
Note that the parameter $q$, through equation (\ref{casca3}), offers a 
quantitative picture of the transition from 
small-scale intermittent, nonextensive, fractal behavior to large-scale 
Gaussian, extensive homogeneity.
For higher values of $n$ and $q=1$, at the top of the cascade, we have
$D_F = 3$. At the bottom of the cascade ($n =1$), using the value
$q_{\eta} \simeq 9/7$, estimated previously, we get $D_F \simeq 2.33$,
a good approximation of the so-called `magic' value of 2.35, often
measured in different experimental contexts$^{24-27}$. This result and
the divergence of the correlation function of the energy 
dissipation in the fluid$^{9}$
\begin{equation}
\label{dis}
\langle \epsilon(x)\epsilon(x+r) \rangle \sim \frac{\langle v_r^6 \rangle}{r^2} = 
\frac{1}{r^2} \left[ \frac{15 \beta^{-2}}{(5 - 3 q)(7 - 5 q)(9 - 7 q)} \right]
\end{equation}
suggest a more stringent bound on $q$, namely $q < 9/7$. If true, this new bound
immediately implies that, in fully developed turbulence, 
relation $(\eta/L)^{\alpha} = 2$
is invariant, regardless the Reynolds number. 

Though only qualitatively, the above picture may be also applied to the 
information cascade. However, since there is nothing equivalent
to viscous damping in the dynamics of speculative markets, the information 
cascade depth is only limited by the minimum time necessary to perform a 
trading transaction.
 
\bigskip

{\bf References}

1. Mantegna, R.N. \& Stanley, H.E. Scaling behaviour in the dynamics of an
economic index. {\it Nature} {\bf 376}, 46-49 (1995).

2. Ghashghaie, S., Breymann, W., Peinke, J., Talkner, P. \& Dodge, Y. 
Turbulent cascades in foreign exchange markets. {\it Nature} {\bf 381}, 
767-770 (1996).

3. Mantegna, R.N. \& Stanley, H.E. Turbulence and financial markets. 
{\it Nature} {\bf 383}, 587-588 (1996).

4. Tsallis, C. Possible generalization of Boltzmann-Gibbs statistics. 
{\it J. Stat. Phys.} {\bf 52}, 479-487 (1988).

5. Tsallis, C., S\'a Barreto, F.C. \& Loh, E.D. Generalization of the Planck
radiation law and application to the cosmic microwave background radiation. 
{\it Phys. Rev. E} {\bf 52}, 1447-1451 (1995).

6. Tsallis, C., Levy, S. V. F., Souza, A. M. C. \& Maynard, R. 
Statistical-mechanical foundation of the ubiquity of L\'evy 
distributions in nature. {\it Phys. Rev. Lett.} {\bf 75}, 3589-3593 (1995).

7. Lyra, M. L. \& Tsallis, C. Nonextensivity and multifractality in 
low-dimensional dissipative systems {\it Phys. Rev. Lett.} {\bf 80}, 53-56 
(1998).

8. Kolmogorov, A. M. A refinement of previous hypotheses concerning the 
local structure of turbulence of a viscous incompressible fluid at
high Reynolds number. {\it J. Fluid Mech.} {\bf 13}, 82-85 (1962).

9. Anselmet, F., Gagne, Y., Hopfinger, E. J. \& Antonia, R. A. High-order 
velocity structure functions in turbulent shear flows. {\it J. Fluid Mech.} 
{\bf 140}, 63-89 (1984).

10. Castaing, B., Gagne, Y. \& Hopfinger, E. J. Velocity probability 
density functions of high Reynolds number turbulence. {\it Physica D} 
{\bf 46}, 177-200 (1990).

11. Benzi, R., Biferale, L., Paladin, G., Vulpiani, A. \& Vergassola, M.
Multifractality in the statistics of the velocity gradients in turbulence.
{\it Phys. Rev. Lett.} {\bf 67}, 2299-2302 (1991).

12. Sreenivasan, K. R. \& Kailasnath, P. An update on the intermittency
exponent in turbulence. {\it Phys. Fluids} {\bf 5}, 512-514 (1992).
  
13. Chabaud et al. Transition towards developed turbulence. 
{\it Phys. Rev. Lett.} {\bf 73}, 3227-3230, (1994). 

14. Praskovsky, A. \& Oncley, S. Measurements of the Kolmogorov constant and
intermittency exponent at very high Reynolds number. {\it Phys. Fluids}
{\bf 6}, 2886-2888 (1994)

15. Vassilicos, J.C. Turbulence and intermittency 
{\it Nature} {\bf 374}, 408-409 (1995).

16. Naert, A., Castaing, B., Chabaud, B., B. H\'ebral \& Peinke, J.
Conditional statistics of velocity fluctuations in turbulence. {\it Physica D} 
{\bf 113}, 73-78 (1998).

17. Taylor, S. J. {\it Math. Fin.} Modeling stochastic volatility: a review
and comparative study. {\bf 4}, 183-204 (1994).

18. Boghossian, B. M., Navier-Stokes equations for generalized thermostatistics.
{\it Braz. J. Phys.} (in the press).

19. Maddox, J., When entropy does not seem extensive. {\it Nature} 
{\bf 365}, 103- (1993).

20. Mandelbrot, B. B. Intermittent turbulence in self-similar cascades: 
divergence of high moments and dimension of the carrier. {\it J. Fluid Mech.} 
{\bf 62}, 331-358 (1974).

21. Parisi, G. \& Frisch, U. in {\it Turbulence and Predictability in
Geophysical Fluid Dynamics and Climatic Dynamics} (ed. Ghil, M., Benzi, R.,
\& Parisi, G.) 84- (North-Holland, Amsterdam, 1985).
  
22. Paladin, G. \& Vulpiani, A. Anomalous scaling laws in multifractal objects.
{\it Phys. Rep.} {\bf 156}, 147-225 (1987).

23. Meneveau, C. \& Sreenivasan, K. R. The multifractal nature of turbulent 
energy dissipation. {\it J. Fluid Mech.} {\bf 224}, 
429-484 (1991).

24. Hentschel, H. G. E. \& Procaccia, I. Relative diffusion in turbulent media:
the fractal dimension of clouds. {\it Phys. Rev. A} {\bf 29}, 1461-1470, (1984). 

25. Sreenivasan, K. R. \& Meneveau, C. Fractal facets of turbulence. 
{\it J. Fluid Mech.} {\bf 173}, 357-386 (1986).
 
26. Sreenivasan, K. R., Ramshankar, R.  \& Meneveau, C. Mixing, entrainment and 
fractal dimensions of surfaces in turbulent flows. {\it Proc. R. Soc. Lond. A}
{\bf 421}, 79-108 (1989).

27. Vassilicos, J. C. \& Hunt, J. C. R. Fractal dimensions and spectra of 
interfaces with application to turbulence. {\it Proc. R. Soc. Lond. A}
{\bf 435}, 505-534 (1991).

\bigskip

{\bf Acknowledgments:} We thank C. Tsallis for helpful discussions. This
work was supported by FAPESP-Brazil and CNPq-Brazil.

\bigskip

\newpage

{\bf Captions}

{\bf Figure 1} (a) Data points: standardized probability distribution $p_q(v_r)$
of velocity differences $v_r(x) = v(x) - v(x+r)$ for spatial scales 
$r = 0.0073L$, $0.0407L$, $0.3036L$, $0.7150L$, with $L/\eta=454$ and $L$ and 
$\eta$ being, respectively, the integral and Kolmogorov scales (see text); data
taken from ref. 2, provided by Chabaud {\it et al.}$^{13}$; Solid lines:
least-squares fits of modified PDF (\ref{pdf}); from top to bottom: $q= 1.26$, 
1.20, 1.11, 1.08; $\beta^- = 0.69$, 0.66, 0.55, 
0.62; $\beta^+ = 0.88$, 0.82, 0.76, 0.70 (for better visibility the curves have 
been vertically shifted with respect to each other).

\bigskip

{\bf Figure 1} (b) Data points: standardized probability distribution $p_q(z_{\tau})$
of price differences $z_{\tau} = z(t) - z(t + \tau)$ for temporal scales 
$\tau = 0.0035\tau_L$, $0.0276\tau_L$, $0.2210\tau_L$, $0.8838\tau_L$, with 
$\tau_L=186265~s$ being the integral scale (see text); data
taken from ref. 2, provided by Olsen \& Associates; Solid lines:
least-squares fits of modified PDF (\ref{pdf}); from top to bottom: $q= 1.35$, 
1.26, 1.16, 1.11; $\beta^- = 1.12$, 0.83, 0.75, 
0.75; $\beta^+ = 0.98$, 0.72, 0.61, 0.77.(for better visibility the curves have 
been vertically shifted with respect to each other).

\bigskip

{\bf Figure 2} Dependence of the parameter $q$ on normalized spatial ($r/L$) and
temporal scales ($\tau/\tau_L$); turbulence data: solid line (model), open squares 
(data, asymmetric adjust), solid squares (data, symmetric adjust); financial data: 
dotted line (model), open triangles (data, asymmetric adjust), solid triangles 
(data, symmetric adjust).

\newpage

\begin{figure}[ppp]
\begin{center}
\mbox{\psfig{file=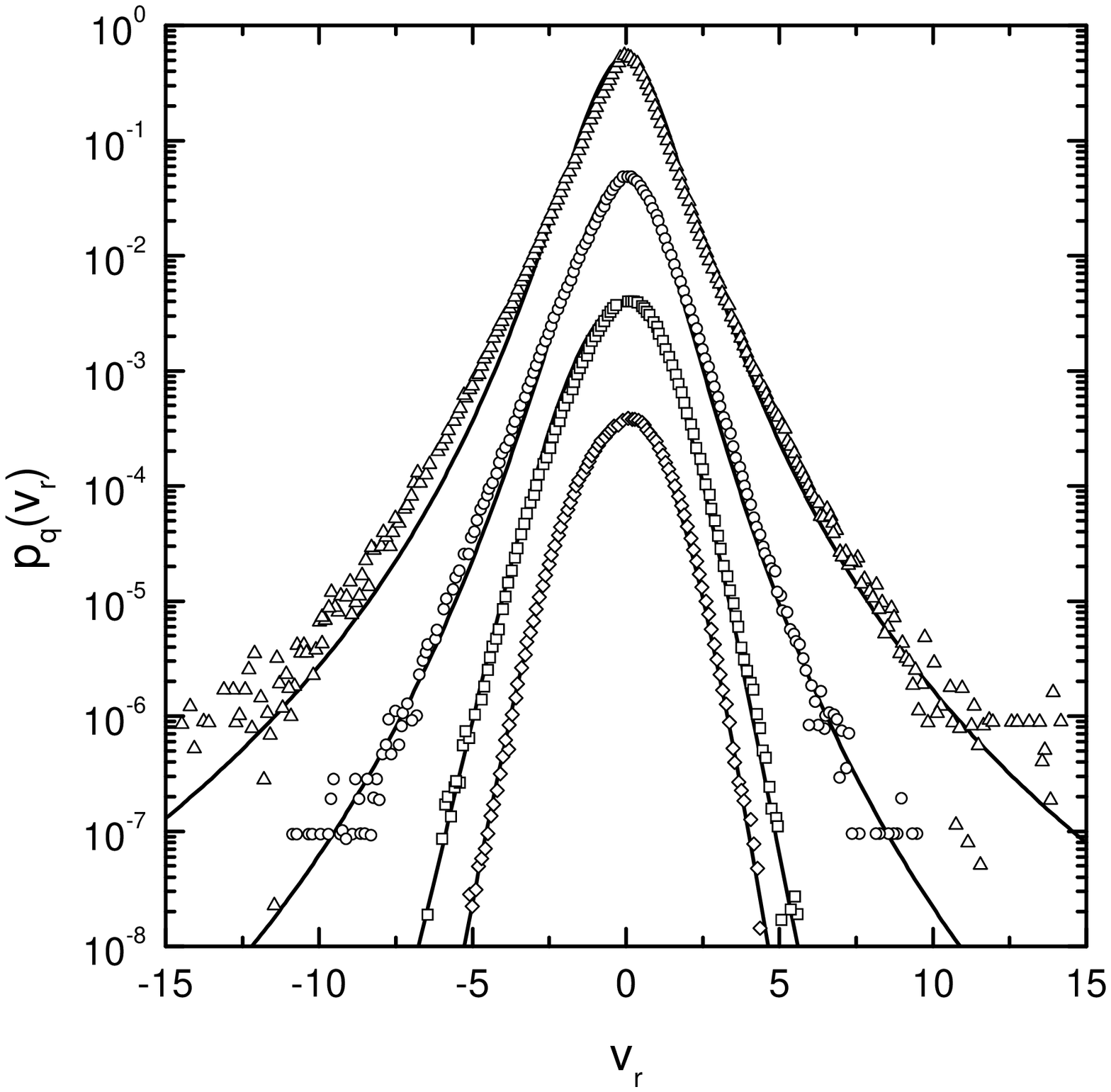,width=14cm}}
\end{center}
\end{figure}

\newpage

\begin{figure}[ppp]
\begin{center}
\mbox{\psfig{file=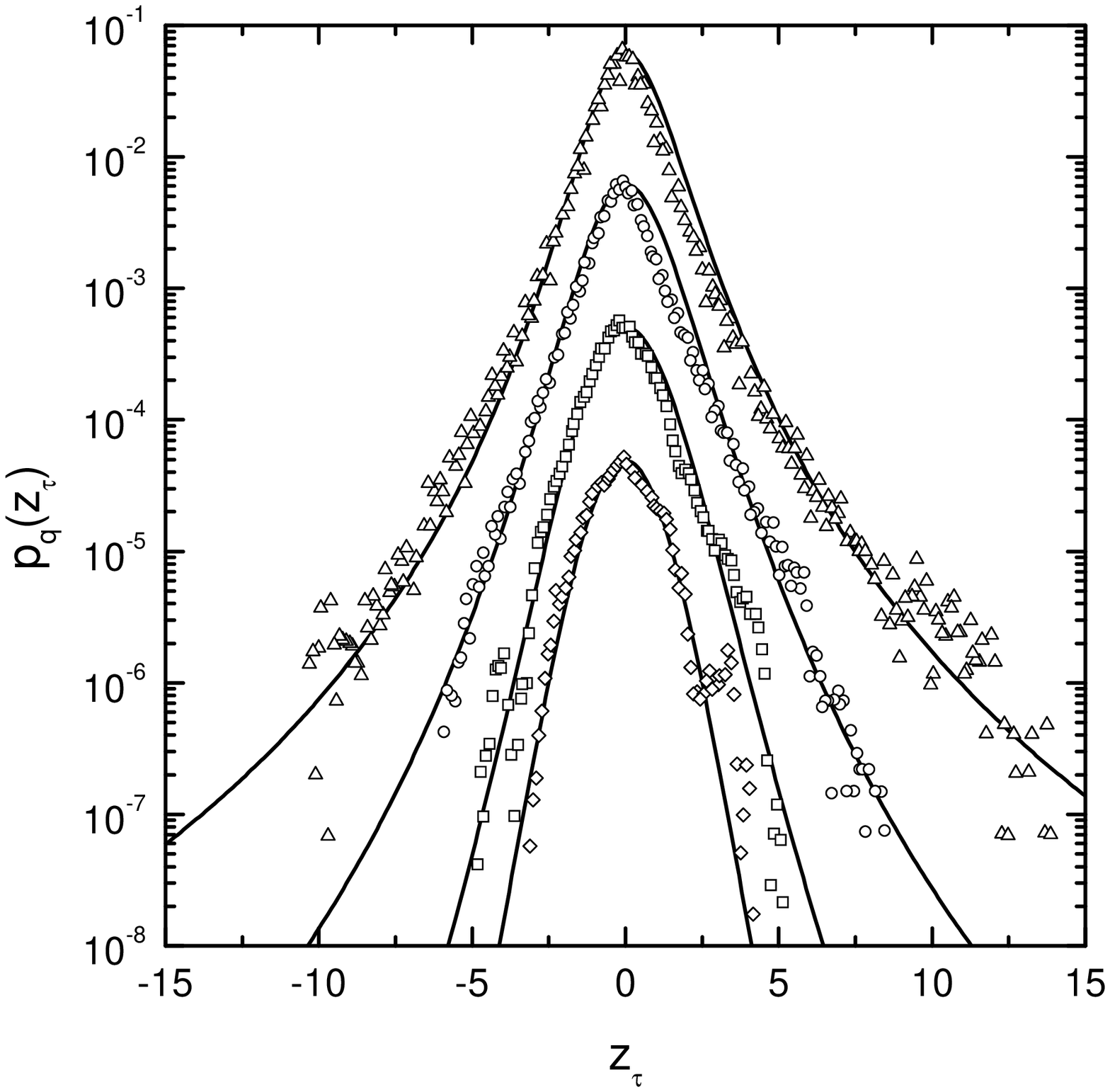,width=14cm}}
\end{center}
\end{figure}

\newpage

\begin{figure}[ppp]
\begin{center}
\mbox{\psfig{file=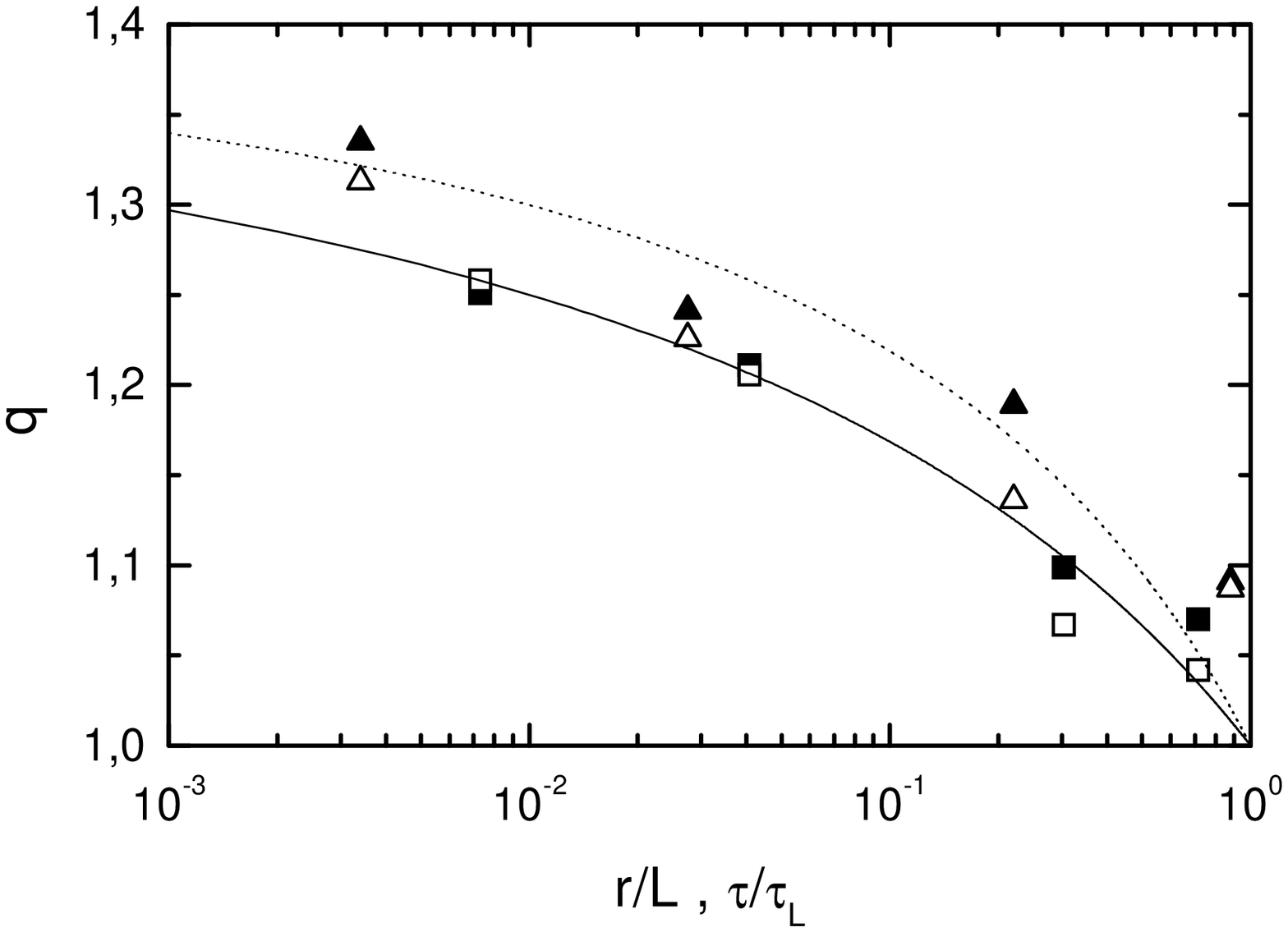,width=14cm}}
\end{center}
\end{figure}

\end{document}